\documentclass[preprintnumbers,amsmath,amssymb,twocolumn,showpacs]{revtex4}
\usepackage{graphicx}% Include figure files
\usepackage{dcolumn}% Align table columns on decimal point
\usepackage{bm}% bold math

\begin{document}

\title{A Single Atom as a Mirror of an Optical Cavity}

\author{G. H\'etet$^{1}$, L. Slodi\v{c}ka$^1$,  M. Hennrich$^1$, and R. Blatt$^{1,2}$}

\affiliation{
$^1$ Institute for Experimental Physics, University of Innsbruck, A-6020 Innsbruck, Austria \\
$^2$ Institute for Quantum Optics and Quantum Information of the
Austrian Academy of Sciences, A-6020 Innsbruck, Austria}

\begin{abstract}
By tightly focussing a laser field onto a single cold ion trapped in front of a far-distant dielectric mirror, we could observe a quantum electrodynamic effect whereby the ion behaves as the optical mirror of a Fabry-P\'erot cavity.
We show that the amplitude of the laser field is significantly altered due to a modification of the electromagnetic mode structure around the atom in a novel regime in which the laser intensity is already changed by the atom alone. We propose a direct application of this system as a quantum memory for single photons.
\end{abstract}
\maketitle

Atom-photon interactions are essential in our understanding
of quantum mechanics. Besides the two processes of absorption and emission of photons, coupling of radiation to atoms raises a number of questions that
are worth investigating for a deeper theoretical and thus interpretational insight. The modification of the vacuum by boundaries is amongst the most fundamental problems in quantum mechanics and is widely investigated experimentally.
In quantum optics, most studies  make use of optical cavities that modify the vacuum-mode density of the field around atoms to change their emission properties \cite{Bru94, Hoo00, Pin00, Kre04}. Another more recent research area investigates the direct coupling of tightly focussed light to atoms in free space, using high numerical aperture elements \cite{Sor07,Son07,Zum08,Ger07,Vam07,Hwa09,Abd10}. There, precise control over the motion of the individual atoms
is crucial to reach the regime of strong atom-light interactions \cite{Teo10}. Recent research in this direction has been performed using cold neutral rubidium atoms \cite{Tey08}, single cold molecules \cite{Wri08}, quantum dots \cite{Vam07}, super-conducting circuits \cite{Abd10} as well as single trapped ions \cite{Sto09,Pir11}. The strong confinement offered by Paul traps, the readily available sideband-cooling techniques, and the ability to perform efficient and deterministic quantum gates \cite{Hae08} make single ions good candidates for such free space quantum communications \cite{Olm09}.

In this Letter, we present a first step towards merging the field of cavity QED with free-space coupling, using an ion trap apparatus. We set up a novel atom-mirror system in which a weak probe field is tightly focussed onto a single trapped ion at the focus of a lens-mirror system. The atomic coupling to the probe is thereby modified by a single mirror in a regime where the probe intensity is already significantly altered by the atom without the mirror.
Furthermore, we show that in the limit of a high numerical-aperture lens, the mirror-induced change in the vacuum-mode density around the single atom can in principle modulate the atom's coupling to the probe, the total spontaneous decay and the Lamb shift, so that the atom behaves as the mirror of a high-finesse cavity.
A measurement of the latter two quantities was in fact performed in \cite{Esc01,Wil03}, by monitoring the excited state population through fluorescence detection. Absorption spectroscopy here enables us to measure the first-order coherence between the driving laser and the back-scattered light and thus to estimate the amplitude of the coherently back-scattered field. Finally, we show that our set-up allows almost full suppression and enhancement by a factor of two of the atomic coupling constant in the probe mode.

\begin{figure}[t]
\centerline{\scalebox{0.4}{\includegraphics{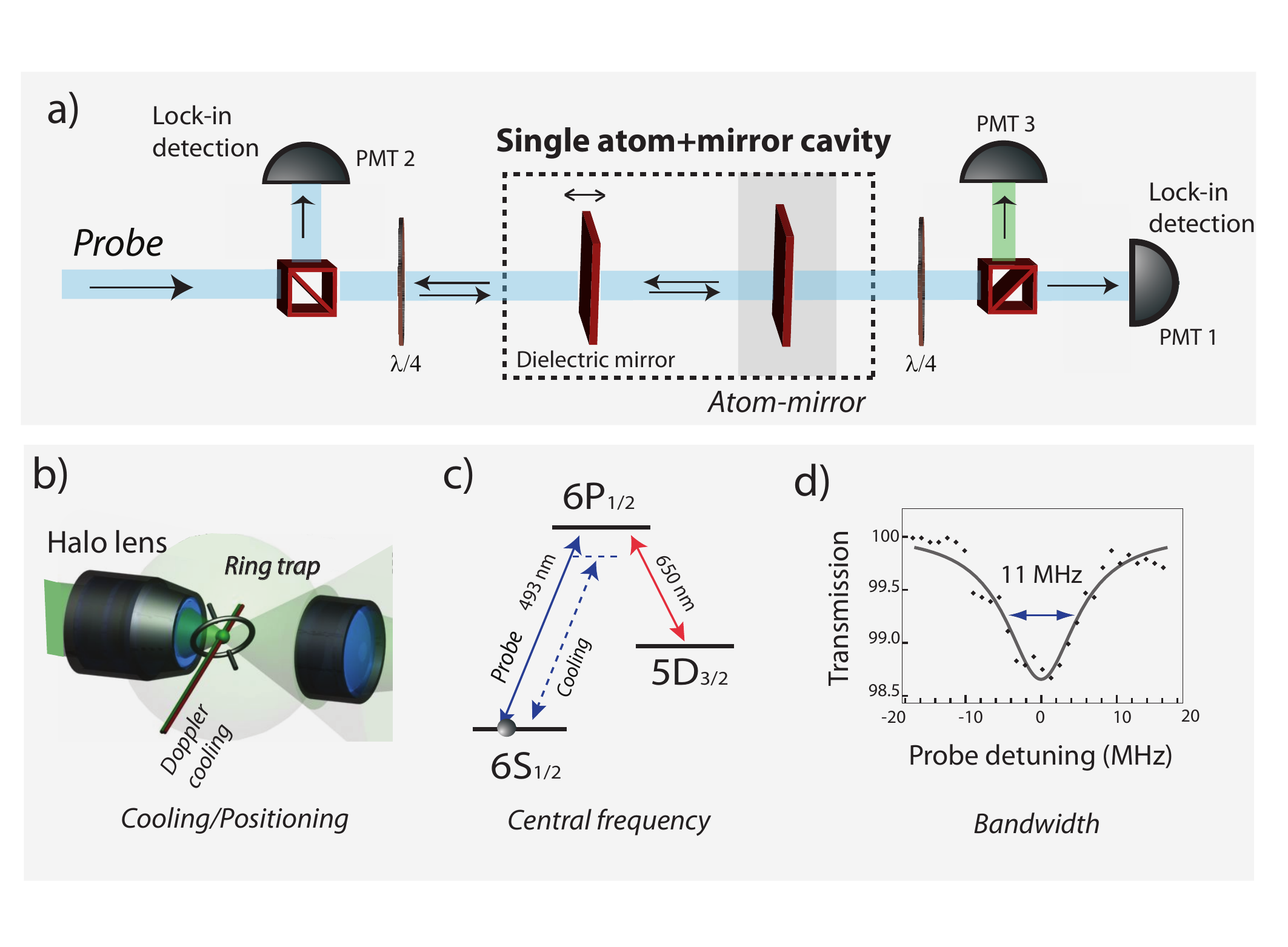}}}
\caption{a) Single ion+mirror set-up. The probe field is coupled to the atom-mirror cavity through the dielectric mirror that is mounted on piezo stages. The intensity of the probe is measured in transmission by PMT1 and in reflection by PMT2. PMT3 is used for measuring the ion fluorescence.  The main properties of the single atom operated as a mirror are shown in  b):  positioning, c): central frequency and d): transmission, as measured without the dielectric mirror. }\label{setup}
\end{figure}

We first consider the single atom as an optical reflector, as depicted in the set-up of Fig.~\ref{setup} a).
Fig. ~\ref{setup} b),~c) and d) show the positioning, central frequency, and transmission bandwidth of the single atom-mirror respectively.
We use a single $^{138}$Ba$^{+}$ ion in a ring Paul trap \cite{Slo10}.
As shown in Fig.~\ref{setup}-c), a narrow-band laser field at 493 nm provides Doppler cooling 50 MHz red detuned from the S$_{1/2}$-P$_{1/2}$ transition, while a laser at 650 nm
recycles the atomic population from the D$_{3/2}$ manifold. The cooling beam intensity is set far below saturation
yet allowing cooling to the Lamb-Dicke regime with a typical final
 population of about $\langle n \rangle\approx 13$.

For extinction of a laser field by the ion in free space,
we use a very weak probe beam resonant with the S$_{1/2}(m_F=+1/2)$-P$_{1/2}(m_F=-1/2)$ transition. As shown in Fig.~\ref{setup} b), the probe beam is overlapped with part of the dipole emission pattern of the ion
using a custom-designed objective with a
numerical aperture (NA) of 0.4. A 1.5\% fraction of the ion's 493 nm
fluorescence together with the transmitted part of the probe beam
is then collected by a microscope objective
and detected by the photomultipliers PMT3 and PMT1, respectively.
Intensity modulation of the
650 nm laser beam, as described in \cite{Slo10}, enables us to efficiently discriminate the fluorescence from the extinction signal.
Fig.~\ref{setup} d) shows the typical Lorentzian dependence of the transmission profile, measured without the dielectric mirror. It shows a width of 11 MHz, and a maximal
extinction of 1.35\%.

In the case of coherent reflection of a laser field by a single atom, the back-scattered field must interfere with the driving laser. To verify this, we construct the system shown in Fig.~\ref{setup} a) by inserting a dielectric mirror 30~cm away from the atom into the
probe path, with a reflectivity $|r|^2=1-|t|^2=99.7\,\%$. We align it so that the ion is re-imaged onto itself and shine the resonant probe through it. Using the Fabry-P\'erot cavity transmissivity, and modeling the atom as a mirror with amplitude reflectivity $2\epsilon$ \cite{Koc94}, one can naively assume that the
intensity transmissivity of the probe reads
\begin{eqnarray}\label{EQ1}
T=\Big|\frac{t(1- 2\epsilon) }{1- 2r\epsilon e^{i\phi_L} }\Big|^2,
\end{eqnarray}
where $\phi_L=2k_LR$, $R$ is the atom-mirror distance and $k_L$ the input probe wavevector.
The finesse $\mathcal{F}=\pi 2\epsilon r/(1-(2\epsilon r)^2)$ of such a cavity-like set-up can in fact be made very large by using a high numerical aperture lens such that $\epsilon\rightarrow 50\,\%$ together with a highly reflective dielectric mirror.
In our experiment, the atom reflectivity is less then 1\% \cite{Slo10}, so the transmitted intensity is well approximated by
\begin{eqnarray}\label{TransSmallEps}
T\approx |t|^2 |1- 2\epsilon+ 2\epsilon r e^{i\phi_L}|^2.
\end{eqnarray}
By tuning the distance between the dielectric mirror and the ion, one would therefore expect a dependence of the transmitted signal on the cavity length, provided that the temporal coherence of the incoming field is preserved upon single-atom reflection.

\begin{figure}[t]
\centerline{\scalebox{0.35}{\includegraphics{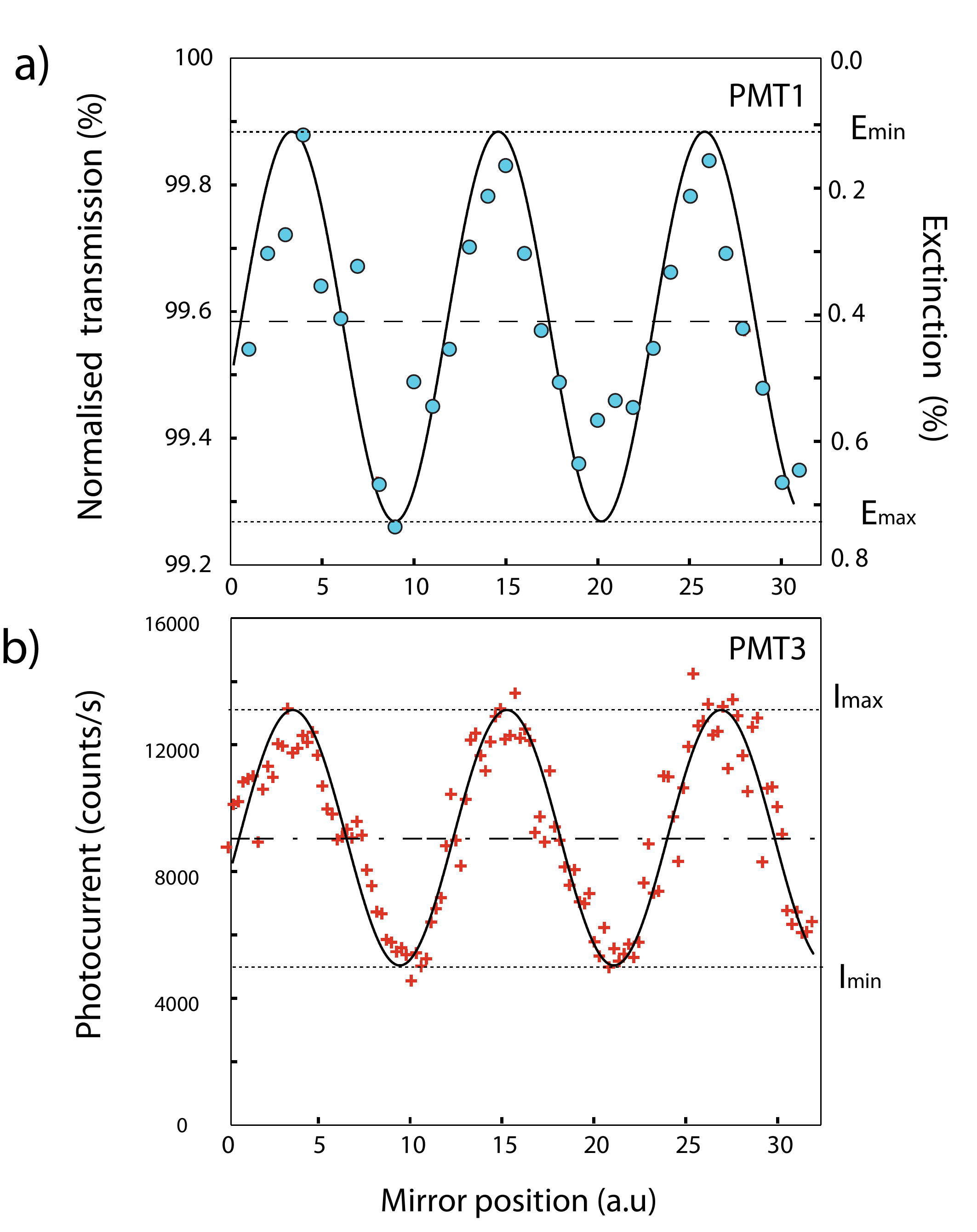}}}
\caption{a) Normalised transmission $T/|t|^2$ of the probe through the single atom-mirror system as a function of the mirror position, with a 99.7 \% reflective dielectric mirror.
The dashed lines shows the transmission of the probe when the mirror is slightly misaligned. The dotted lines show the minimum and maximum extinction values used for estimating the contrast $V'$.
b) The single photon interference fringe measured on PMT3.
Solid lines are the sinusoidal fits to the data. The dotted lines show the minimum and maximum photocurrent values used for estimating the contrast $V$. Error bars are on the order of 0.05\% for all the points.}\label{front}
\end{figure}

\begin{figure}[t]
\centerline{\scalebox{0.35}{\includegraphics{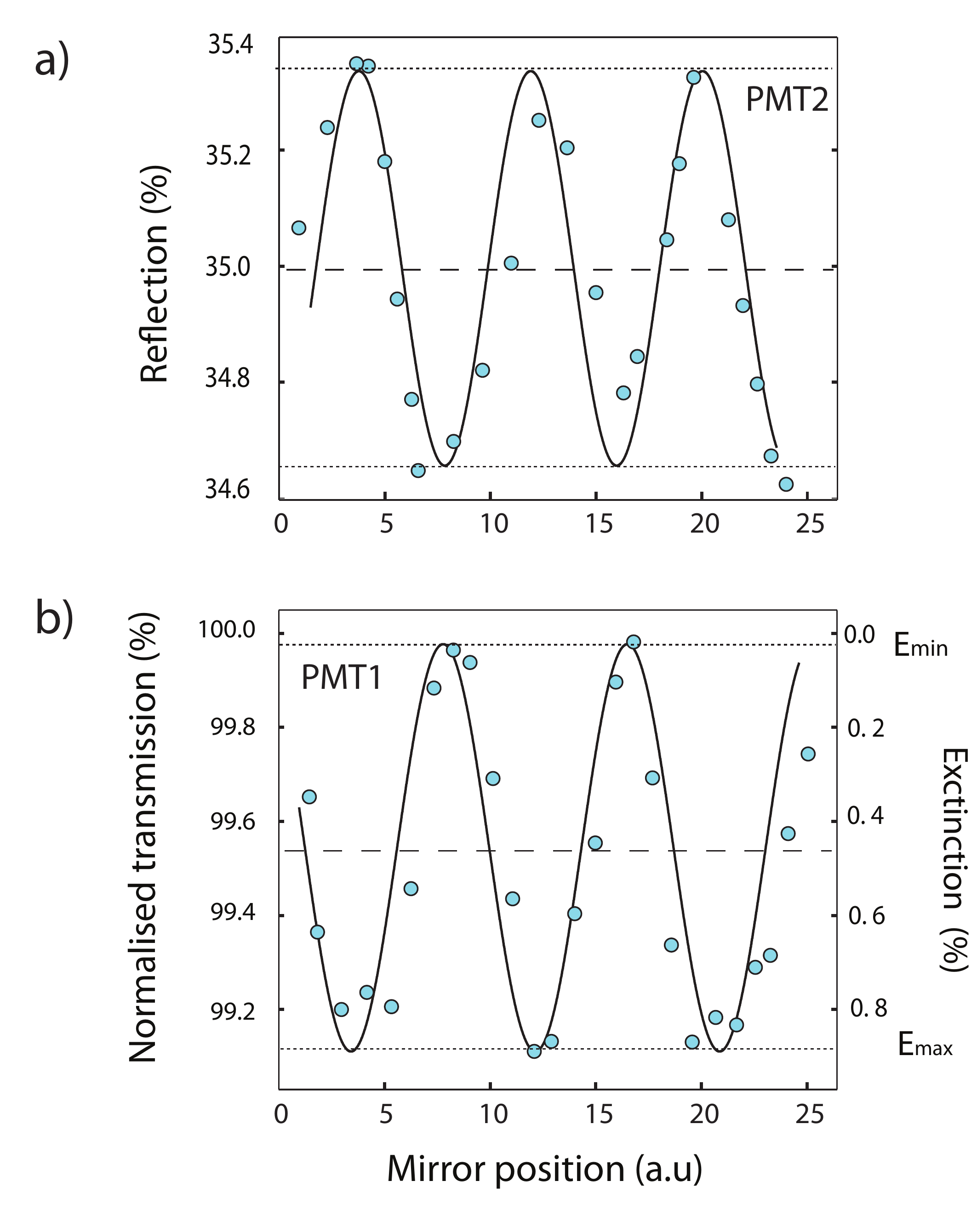}}}
\caption{a) Intensity of the probe reflected off the cavity, normalized to the probe intensity without ion, and using a 75\% reflective dielectric mirror.
b) shows the transmission of the probe through the atom-mirror system as a function of the mirror position, normalized to the mirror transmissivity. The dashed line in a) shows the reflection of the probe without the ion and in b) the transmission with the mirror misaligned from the ion.}\label{back}
\end{figure}

The operation of our ion-mirror system is shown in Fig.~\ref{front} a). As the mirror position is scanned, we indeed observed clear sinusoidal oscillations of the power detected in PMT1 on a wavelength scale. These results reveal that the elastic back-scattered field is interfering with the transmitted probe, and that the ion is very well within the Lamb-Dicke regime.
Fig.~\ref{front} b) shows the fluorescence rate measured at PMT3 for the same experimental conditions but with the probe field blocked. The intensity change of the fluorescence rate is the result of the self-interference of single photons, which can be expressed as $I=I_0(1+V\cos(\phi_L))$ \cite{Esc01,Dor02}.
With our ion-mirror distance (30~cm), the interference contrast $V$ is mostly limited by residual aberrations of the imaging optics and atomic motion \cite{Esc01}. As predicted by the formula for the transmission $T$ (Eq. \ref{TransSmallEps}), the two signals in Fig.~\ref{front} a) and Fig.~\ref{front} b) oscillate perfectly in phase. The oscillations are however observed with a lower contrast than for the extinction coefficient (defined as $E=1-T/|t|^2$, and plotted on the right axis). As we will show, this pronounced difference stems from an aberration-free dependence of the extinction contrast.
We then perform another experiment in which we replace the high reflectivity mirror by a 25/75\% mirror. The results are shown in Fig.~\ref{back} a) and b) where
we simultaneously recorded the reflected and transmitted powers measured on PMT2 and PMT1 respectively. With this mirror reflectivity, we are able to measure the change of the probe power being reflected off the cavity, which we found to be exactly out of phase with the transmitted signal, as is predicted for a Fabry-P\'erot cavity response. We note that, here again, an unexpectedly large extinction contrast is observed.

The contrast $V'=(E_{\rm max}-E_{\rm min})/(E_{\rm max}+E_{\rm min})$ of the ion+mirror cavity extinction plotted in Fig.~\ref{front} a) and \ref{back} b), and the ion's single photon interference contrast in Fig.~\ref{front} b) clearly differ.
%Such a feature does not appear in the single photon self-interference of the ion's fluorescence because the directly emitted field and the retro-reflected field do not travel through the same lens.
To understand this effect, we will consider the influence of aberrations, by including a phase shift to each of the contributing amplitudes of the transmitted field at various points on the lens.
As shown in the supplementary material, the transmissivity of the probe, in the limit of a high dielectric mirror reflectivity, is then
%We change the input amplitude $E_{\rm in}\rightarrow E_{\rm in}e^{i\phi'(\vec{r}_l)}$
%so that the input probe gets an $\vec{r}_l$-dependent phase shift when it goes through the point $\vec{r}_l$ on the lens surface. We replace $\epsilon$ by $ \epsilon' $ ($\epsilon>\epsilon'$) to account for a corresponding decrease of the field amplitude at the focus.
%Finally, we make the substitution $2\epsilon e^{i\phi_L} E_{\rm in}\rightarrow 2\epsilon' e^{i\phi_L}e^{2 i\phi'(\vec{r}_l)} E_{\rm in}$, where the reflected scattered field gets two times the $\vec{r}_l$-dependent phase shift.
%Without abberations, the contribution to the probe transmission consists of three terms :
%\begin{eqnarray}
%%T_{\rm out}=|1- 2\epsilon- 2\epsilon e^{i\phi}|^2\approx 1-4\epsilon-4\epsilon\cos(\phi)
%%\end{eqnarray}
%We reach $T \approx |t|^2(1-4\epsilon'\cos\phi(\vec{r}_l)-4\epsilon'\cos(\phi(\vec{r}_l)+\phi_L))$, with a high reflectivity mirror,
%%\begin{eqnarray}
%%T&=&|e^{i\phi'(\vec{r}_l)}- 2\epsilon'- 2\epsilon' e^{i\phi}e^{2 i\phi'(\vec{r}_l)}|^2\\
%%&\approx& 1-4\epsilon\cos\phi(\vec{r}_l)-4\epsilon\cos(\phi(\vec{r}_l)+\phi)
%%\end{eqnarray}
%which, averaged over the lens surface with a corrugation pitch smaller than the optical wavelength $\lambda$, gives the normalized transmission
\begin{eqnarray}
T \approx |t|^2(1-4\overline{\epsilon}(1-\cos(\phi_L)).
\end{eqnarray}
Here $\overline{\epsilon}=\epsilon' J_0(\eta)$, $J_0(\eta)$ is the first order Bessel function of the first kind and $\eta=2\pi \sigma_{\rm ab} /\lambda$ where $\sigma_{\rm ab}$ is the root mean square amplitude of the aberrations and $\lambda$ is the optical wavelength.
We then obtain the normalized extinction plotted Fig.~\ref{front}-a) to be
\begin{eqnarray}\label{EQ6}
E = 4\overline{\epsilon}(1-\cos(\phi_L)).
\end{eqnarray}
The contrast of $E$ is then free of the aberrations that one could expect to play a role.
When making the same substitutions in the formula for the single photon interference that we observed in Fig.~\ref{front}-b), one however gets to the intensity
\begin{eqnarray}
 I=I_0 (1+J_0(\eta)\cos(\phi_L)),
\end{eqnarray}
which shows a direct dependence on the aberrations. The two intensities that contribute to the extinction $E$ in fact arise from an interference between the input and the scattered amplitudes that carry the same global phase shifts. This explains the larger contrast measured Fig.~\ref{front}-a) and Fig.~\ref{back}-b) over Fig.~\ref{front}-b). This observation will be precious for precise characterization and control of the tight focussing of optical fields onto single trapped particles.

We now investigate whether the naive Fabry-P\'erot interpretation that we used to describe our results is valid. One could indeed wonder how the modification of the quantum vacuum around the atom affects our results. It is clear that the dielectric mirror imposes new boundary conditions that will change the vacuum mode density close to the atom, but it is less obvious how much it will contribute to the probe intensity changes that we observe in this experiment. One can in fact show (see supplementary material) that solving the multimode Heisenberg equations in a time-dependent perturbation theory gives
 \begin{eqnarray}\label{EQ3}
T=|t|^2\Big|1-\frac{2g_{\epsilon}\overline{g}^\ast}{\tilde{\gamma}+i\tilde{\Delta}}\Big|^2,
\end{eqnarray}
assuming the input probe to be resonant with the atomic transition.
Here, $g_{\epsilon}$ denote the atomic coupling strength in the probe mode,
$\overline{g}$ is the mean coupling to all the modes, $\tilde{\gamma}$ and $\tilde{\Delta}$ are the decay and level shifts modified by the presence of the mirror. Their expressions can be evaluated using the appropriate spatial mode function for this system \cite{het10} and we can then show that
\begin{eqnarray}\label{EQ5}
\frac{g_{\epsilon}\overline{g}^\ast}{\tilde{\gamma}+i\tilde{\Delta}}= \frac{\epsilon(1- r e^{i\phi_L}) }{1- 2r\epsilon e^{i\phi_L} }.
\end{eqnarray}
After combining this relation with Eq. \ref{EQ3} we obtain the same transmissivity as was obtained by modeling the atom as a mirror with reflectivity $2\epsilon$ (Eq. \ref{EQ1}). Interestingly, the QED calculations yield the same mathematical results as the direct Fabry-P\'erot calculation.

In this QED approach, it was not necessary to invoke multiple reflections off the atom for the Fabry-P\'erot like transmission to appear. The transmission of the probe through the single atom+mirror system is mathematically equivalent to a cavity, but the origin of the peaked transmission profile can be interpreted as a line-narrowing effect due to the QED-induced changes of the spontaneous emission rate and level shift.
In our experiment, we observed a change of the coupling between the atom
  and the probe mode, due to the modification of the mode density induced by the mirror. Deviations from the sinusoidal shape due to line narrowing would be visible for a lens covering a solid angle of more than 10\%. We note that, with this interpretation, the aberration-free dependence of the extinction contrast is analogous to an almost complete cancelation and enhancement by a factor of two of the atomic coupling constant in the probe mode.

We foresee a direct application of our system.
In discrete variable quantum communications, and specifically for quantum repeater architectures, single photons must be stored and released from stationary qubits \cite{Cir97,Pin08} to prevent the unavoidable losses in optical fibers \cite{Bri98}. The required efficient coupling between a single photon and a single atom can be obtained through the use of a high finesse cavity \cite{Cir97,Spe11}, or parabolic mirrors for mode-matching the incoming field with the whole atomic dipole field \cite{Lin07}. Our single atom-mirror set-up is an attractive alternative solution for full absorption of a single photon. In such a scenario, the retro-reflection of the back-scattered field by the mirror mediates the required interference effect so that the excitation probability of the atom can reach more than 50\% \cite{Sto09,Wan10}.
However, unlike standard lossless mirrors, the ion will fully reflect the light back into the probe mode only for $\epsilon=50\,\%$ so in the realistic NA case, the scattered field is emitted into almost $4\pi$ steradian.
Since impedance matching is here not immediately fulfilled, in order to attain a steady state transmission of the optical field through such a system, one can optically pump a fraction of atomic population to another state to match the input mirror reflectivity. Implementing a dynamic coherent transfer of population \cite{Ber98,Pin08} to another metastable ground state will furthermore allow efficient and long lived quantum storage of a single photon pulse in the atom. Alternatively, one could ramp the mirror position from the anti-node to the node of the standing wave so as to match the incoming photon's temporal profile to the ion-mirror system and store the photon in the long lived atomic excited state \cite{het10}.
Although the present results are obtained in the elastic scattering regime with two levels, our experimental results may be seen as the first tests of such a new single atom-photon interface.

In conclusion, we successfully observed the operation of a single atom as an optical mirror of a Fabry-P\'erot-like cavity. Our investigations are performed in a novel regime where a significant fraction of the power of a probe field can be affected by the atom in free space. This allows us to realize an experiment in which both the properties of an atom as a reflector and the modification of the atomic coupling constant can play a role.
Although a simple cavity interpretation lends itself naturally to a description of our experiment, a more general QED formulation should be preferred for an unambiguous discrimination of the involved mechanisms. Interestingly, for our experimental parameters (weak excitation, small atom-mirror distances), we found that both interpretations are equivalent. Besides the appealing quantum memory application that we presented above, our set-up has a number of other realistic prospects. It will, for instance, be a useful tool for operating the ion as an optical switch, similar to the single atom transistors using EIT implemented in \cite{Slo10, Abd10,Mue10,Kam10} or using a population inversion \cite{Hwa09}.

We thank T. Northup and N. R\"ock for carefully checking the manuscript and A. Gl\"atzle for useful discussions.
This work has been partially supported by the Austrian
Science Fund FWF (SFB FoQuS), by the European Union
(ERC advanced grant CRYTERION) and by the Institut fur
Quanteninformation GmbH. G. H. acknowledges support
by a Marie Curie Intra-European Action of the European
Union.

\newpage

\section*{Supplementary materials}

\subsection{The role of aberrations}

We give here a more detailed account about the effects of aberrations.

The modulation of the fluorescence rate in Fig. 2 b),
is due to a self-interference of the single photons emitted by the atom \cite{Esc01,Dor02}. The intensity change of the fluorescence can be expressed as $I=I_0(1+V\cos(\phi_L))$. With our ion-mirror distance (30~cm), the interference contrast $V=(I_{\rm max}-I_{\rm min})/(I_{\rm max}+I_{\rm min})$ is mostly limited by residual aberrations of the imaging optics and atomic motion \cite{Esc01}. As predicted by the formulas for the transmission $T$ (Eq. 2) and fluorescence intensity $I$, the two signals in Fig.~2 a) and Fig.~2 b) oscillate in phase. We note however, that the contrast $V'=(E_{\rm max}-E_{\rm min})/(E_{\rm max}+E_{\rm min})$ of the ion+mirror cavity extinction in Fig.~2 a) and defined on the right axis, and the ion's fluorescence self-interference contrast in Fig.~2 b) clearly differ.

To understand this effect, we will consider the influence of aberrations by including a random phase shift to each of the contributing amplitudes of the transmitted field. We change the input field amplitude $E_{\rm in}\rightarrow E_{\rm in}e^{i\phi'(\vec{r}_l)}$
so that the input probe gets an $\vec{r}_l$-dependent phase shift when it goes through the point $\vec{r}_l$ on the lens surface. We replace $\epsilon$ by $ \epsilon' $ ($\epsilon>\epsilon'$) to account for a decrease of the field amplitude at the focus due to these same phase shifts.
Finally, we make the substitution $2\epsilon e^{i\phi_L} E_{\rm in}\rightarrow 2\epsilon' e^{i\phi_L}e^{2 i\phi'(\vec{r}_l)} E_{\rm in}$, where the reflected scattered field gets two times the $\vec{r}_l$-dependent phase shift.
%Without abberations, the contribution to the probe transmission consists of three terms :
%\begin{eqnarray}
%T_{\rm out}=|1- 2\epsilon- 2\epsilon e^{i\phi}|^2\approx 1-4\epsilon-4\epsilon\cos(\phi)
%\end{eqnarray}
For simplicity, we here consider the case of a high mirror reflectivity.
We then obtain $T(\vec{r}_l) \approx |t|^2(1-4\epsilon'\cos\phi(\vec{r}_l)-4\epsilon'\cos(\phi(\vec{r}_l)+\phi_L))$,
%\begin{eqnarray}
%T&=&|e^{i\phi'(\vec{r}_l)}- 2\epsilon'- 2\epsilon' e^{i\phi}e^{2 i\phi'(\vec{r}_l)}|^2\\
%&\approx& 1-4\epsilon\cos\phi(\vec{r}_l)-4\epsilon\cos(\phi(\vec{r}_l)+\phi)
%\end{eqnarray}
which, averaged over the lens surface with a corrugation pitch smaller than the optical wavelength $\lambda$, gives the total transmission
\begin{eqnarray}
T \approx |t|^2(1-4\overline{\epsilon}(1-\cos(\phi_L)).
\end{eqnarray}
Here $\overline{\epsilon}=\epsilon' J_0(\eta)$, $J_0(\eta)$ is the first order Bessel function of the first kind and $\eta=2\pi \sigma_{\rm ab} /\lambda$ where $\sigma_{\rm ab}$ is the root mean square amplitude of the aberrations.
We then obtain the normalized extinction plotted in Fig.~2-a) to be
\begin{eqnarray}
E = 4\overline{\epsilon}(1-\cos(\phi_L)).
\end{eqnarray}
When making the same substitutions in the case of the single photon interference that we observed in Fig.~2-b), one gets the intensity
\begin{eqnarray}
 I=I_0 (1+J_0(\eta)\cos(\phi_L)),
\end{eqnarray}
The contrast of $I$ and $E$ will therefore differ, since the contrast of the single photon interference fringes depends directly on the aberrations. The two intensities that contribute to the extinction $E$ arise from an interference between the input and the scattered amplitudes that carry the same global phase shifts, which is the reason for the larger contrast observed Fig.~2-a) and Fig.~3-b) over Fig.~2-b).
 %The interference terms between the probe field and the scattered light are $|e^{i\phi(\vec{r}_l)}-2\epsilon|^2$, and the probe and retro-reflected scattered light $|e^{i\phi(\vec{r}_l)}-2\epsilon e^{2i\phi(\vec{r}_l)}|^2$ which are the same.

To find out how much of the single photon-interference contrast is limited by abberations, we used an independent estimation method by coupling the fluorescence into a single mode fiber. A contrast close to 90\% was observed in this case, which shows that most of the self-interference contrast is in fact limited by aberrations.
$J_0(\eta)$ is therefore about 30\%-40\% in our measurements.
We note that atomic motion would also give aberration-like effects,
however a separation of these two sources of error is not straightforward and will be the subject of further studies.

%The interference terms between the probe field and the scattered light are $|e^{i\phi(\vec{r}_l)}-2\epsilon|^2$, and the probe and retro-reflected scattered light $|e^{i\phi(\vec{r}_l)}-2\epsilon e^{2i\phi(\vec{r}_l)}|^2$ which are the same.

%A direct measurement of the elastic back-reflected field was also proven to be a difficult task since its intensity is only proportional to $\epsilon^2$, as opposed to $\epsilon$ for the extinct part of the probe. This is because, in the latter case, the forward scattered field beats with the probe field \cite{Alj10}. Interfering the back-scattered field with the probe in such atom-mirror system therefore allows its efficient measurement.

\subsection{Quantum electrodynamics calculations}

Here, we present the calculations showing the equivalence between the QED and scattering approaches. The simple cavity description that we used initially is intuitive but the extent to which the ion behaves as a mirror in the presence of other boundaries is an open question. To make sure our cavity interpretation is valid, the QED effects including the atomic dynamics will either have to give the same mathematical results or be negligible.
We will show that treating the atom as a mirror or doing the QED calculations with a resonant atom light coupling yields the same results, but with a different physical interpretation.

The free part of the light-atom Hamiltonian is
\begin{eqnarray}
{H}_0= \frac{1}{2}\hbar\omega_{0}{\sigma}_z + \sum_\mu \hbar\omega_{{\mu}}\Big[{a}_{{\mu}}^{\dagger}{a}_{{\mu}}+\frac{1}{2}\Big],
\end{eqnarray}
where ${\sigma}_z=|e\rangle \langle e|-|g\rangle \langle g|$, is the difference of population between the excited and ground states $|e\rangle$ and $|g\rangle$ respectively. ${a}^{\dagger}_{{\mu}}$ denotes the creation operator for a photon in a mode $\mu$ of the reservoir,
$\omega_0$ is the atomic transition frequency, $\omega_\mu$ is the frequency of the optical mode $\mu$. The interaction Hamiltonian in the Coulomb gauge and in the electric dipole approximation is
\begin{eqnarray}
{H}_{\rm int}=-\frac{e}{m c}\vec{A}(\vec{r},t)\cdot \vec{p},
\end{eqnarray}
where $\vec{p}$ is the momentum operator of the atomic electron, $\vec{r}$ its position and $m$ its mass. $\vec{p}=m \omega_0\vec{d}/e\times{\sigma}_{y}$, where ${\sigma}_{y}=i({\sigma}-{\sigma}^\dagger)$, $\vec{d}$ is the electric dipole matrix element of the two level atom, $\vec{A}$ is the vector potential, and $\sigma=|g\rangle \langle e|$. Since we are interested in the modification of real photon processes when the mirror is far, we did not include the $A^2$ contribution in the Hamiltonian, and will make a rotating wave approximation. We decompose $\vec{A}$ over a complete mode basis $\vec{e}_{\mu}$ as
\begin{eqnarray}
\vec{A}(\vec{r},t)&=& \sum_\mu~ \sqrt{\frac{2\pi \hbar c^2}{\omega_\mu}}~\vec{e}_{\mu}(\vec{r})~{a}_{\mu}(t)+h.c.,
\end{eqnarray}
where the sum is taken over all normalized eigenfunctions $\vec{e}_{\mu}$ of the Helmholtz equation.

It is shown in \cite{het10} that after making a spherical decomposition for the far-field modes, one can decouple the longitudinal and angular integration over the set of mode functions, so that $\sum_\mu\rightarrow \Lambda/\pi\int dk\sum_{l,m}$, where $\Lambda$ is the radius of the quantization sphere, and $(l,m)$ are the quantum numbers for the angular momentum and spin respectively.
Discarding polarization, and assuming the atom to be at the focus, the mode function can be written as
\begin{eqnarray}
e_\mu(\vec{r}=0)= b_l i^l T_{l,m,k},
\end{eqnarray}
with
\begin{eqnarray}\label{ModeFunc}
T_{l,m,k}&=&2i\int_{2\pi} \frac{d\vec{\Omega}}{4\pi} \Big[(1- r_M e^{i\phi})+(-1)^l t_M  \Big]Y_{l,m}.
\end{eqnarray}
where $|r_M|^2=1-|t_M|^2$ is the mirror reflectivity (which is set to zero when the mirror is absent at angles $\vec{\Omega}$), $\phi=2 k R$ where $R$ is the distance between the atom and the mirror, and $b_l$ is the amplitude of the mode $l$. $Y_{l,m}$ are the spherical harmonics.
The total field at the focus is then a superposition of a standing wave (first term in Eq.~\ref{ModeFunc}) and a wave that is coming either through the mirror or from free space (second term).

Let us now write the coupling strength of the atom to the mode $\mu$ as
\begin{eqnarray}
g_{\mu}=C e_\mu,
\end{eqnarray}
where
\begin{eqnarray}
C=-i\omega_0\sqrt{\frac{2\pi}{\hbar \omega_\mu}} d.
\end{eqnarray}
The normally ordered Heisenberg equations for the time-dependent optical modes $\mu$ and atomic operators read
\begin{eqnarray}\label{ME}
\Big(\frac{\partial}{\partial t}+i\Delta_{\mu}\Big) {a}_{\mu}&=&i g^{\ast}_\mu~{\sigma},\\\label{BE}
\Big(\frac{\partial}{\partial t}+i\Delta\Big) {\sigma}&=&-i\sum_\mu g_\mu {\sigma}_z {a}_{\mu},\\
\frac{\partial}{\partial t} {\sigma}_z&=&-2i\sum_\mu (g_\mu {\sigma}^{\dagger}{a}_{\mu}+g^{\ast}_\mu {a}^{\dagger}_{\mu}{\sigma})
\end{eqnarray}
where we moved into a frame at the frequency $\Delta-\omega_0$, where $\Delta=\tilde{\Delta}+\Delta_{AC}$, and introduced $\Delta_{\mu}=\delta_\mu+\Delta$ where $\delta_\mu=\omega_\mu-\omega_0$.
$\tilde{\Delta}$ is the modified Lamb shift and $\Delta_{AC}$ the AC Stark-shift induced by the probe field, the expression of which will be given after solving the Heisenberg equations.

Integration of the equation for the optical field from $t_0$ to $t$ yields
\begin{eqnarray}\label{light}
a_\mu(t)&=&a_{\mu}(t_0)e^{-i\Delta_\mu(t-t_0)}\nonumber\\
&+&ig_\mu^\ast\int_{t_0}^t dt' e^{-i\Delta_\mu(t-t')}\sigma(t').
\end{eqnarray}
%The `mirror' field is the sum of the input field and the scattered field from the atom.
Summing this equation over all $\mu$ gives
\begin{eqnarray}
\sum_\mu  a_\mu(t)=E_c(t)+i\overline{g}^\ast\sigma(t),
\end{eqnarray}
where we made a Markov approximation $\sigma(t')\approx \sigma(t)e^{-i\Delta(t-t')}$.
We introduced
\begin{eqnarray}\label{gave}
\overline{g}=C\sum_{\mu} b_l i^l T_{l,m,k}\int_{t_0}^t dt' e^{-i\delta_\mu(t-t')}.
\end{eqnarray}
We also define
\begin{eqnarray}
E_c(t)=\sum_\mu a_{\mu}(t_0)e^{-i\Delta_\mu(t-t_0)},
\end{eqnarray}
the ``intra-cavity" mode at $t=t_0$, which is directly related to the input probe via
\begin{eqnarray}
E_c(t)=t_M E_{\rm in}(t)+r_M E_{\nu}(t).
\end{eqnarray}
$E_{\nu}(t)$ is the electric field corresponding to the mode $\mu$, with zero expectation value in the vacuum state.

One can do the same integration, for times $t<t_1$, as is done in \cite{Gar85}, and after solving the time-reversed Heisenberg equation, one obtains
\begin{eqnarray}
\sum_\mu a_\mu(t)=E_{\rm out}(t)-i\overline{g}^\ast\sigma(t),
\end{eqnarray}
where
\begin{eqnarray}
E_{\rm out}(t)=\sum_\mu a_{\mu}(t_1)e^{i\Delta_\mu(t-t_1)}.
\end{eqnarray}
Grouping together the two results, one finds
\begin{eqnarray}\label{inpout}
E_{\rm out}(t)=E_{c}(t)+2i\overline{g}^\ast\sigma(t),
\end{eqnarray}
relating the cavity and output fields through the coupling with the atom.

Let us now consider the atomic response to the optical field.
We first solve the equation for ${\sigma}_z(t)$ to first order in the atom-light coupling and insert it in Eq.~(\ref{BE}). One then finds
\begin{eqnarray}\label{SS45}
\Big(\frac{\partial}{\partial t}+i\Delta\Big) \sigma(t)
&=&-(\tilde{\gamma}-i(\tilde{\Delta}+\Delta_{AC})) \sigma(t) \nonumber\\
&-& ig_{\epsilon}  \sigma_z(0)  E_c(t) ,
\end{eqnarray}
where we replaced ${a}_{\mu}(t)$ by Eq.~(\ref{light}), and used again the approximation $\sigma(t')\approx \sigma(t)e^{-i\Delta(t-t')}$.
Here $g_{\epsilon}$ is the coupling strength of the probe modes coupled via the lens. It is given by
\begin{eqnarray}\label{geps}
g_{\epsilon}=C \sum_{l,m} b_l i^l T_{l,m,k_L},
\end{eqnarray}
where
\begin{eqnarray}
T_{l,m,k_L}&=&2i\int_{\epsilon} \frac{d\vec{\Omega}}{4\pi} \Big[(1- r_M e^{i\phi_L})+(-1)^l t_M  \Big]Y_{l,m}.
\end{eqnarray}
We also defined the modified spontaneous emission rate and level shifts as
\begin{eqnarray}
\tilde{\gamma}&=&\sum_\mu |g_\mu|^2 \delta(\omega_\mu-\omega_0),\\
\tilde{\Delta}&=&\sum_\mu |g_\mu|^2 P\Big[\frac{1}{\omega_\mu-\omega_0}\Big],\\
\Delta_{AC}&=&\sum_\mu |g_\mu|^2 \int_{t_0}^t dt' e^{-i\delta_\mu(t-t')} n_\mu(0),
\end{eqnarray}
respectively, all of which are modified due to the mirror back-action \cite{Dor02,het10}.
Here $ {n}_\mu(0) $ is the initial number of photons in the mode $\mu$, that is the number of photons in the probe mode. One can write $ {n}_\mu(0) = {n}_{k_L} $ where $k_L$ is the probe wavevector.

%This yields
%\begin{eqnarray}
%T^{0}_{l,m}=2\epsilon i t e^{i k_0 R}\cos(k_0 R)Y_{l,m},
%\end{eqnarray}
%when the input field covers $\epsilon$ and with the atom at the focus of the lens, so that
%\begin{eqnarray}
%g_{k_0}=C 2\epsilon t e^{i k_0 R}\cos(k_0 R).
%\end{eqnarray}
%The same considerations give
%\begin{eqnarray}
%\overline{g}_{k}=C (1-2\epsilon r e^{i k_0 R}\cos(k_0 R)).
%\end{eqnarray}
Integrating Eq. (\ref{SS45}) yields
 \begin{eqnarray}\label{SSatom}
\sigma(t)&=&-i g_{\epsilon}\sigma_z(0)E_c(t) \int_0^t dt' e^{-(\tilde{\gamma}+i(\Delta+\delta_{k_L}))(t-t')}.
\end{eqnarray}
Combining Eq.~(\ref{SSatom}) and  Eq.~(\ref{inpout}), and assuming the probe to be weak enough so that $\Delta_{AC}\ll\tilde{\gamma}$ , one finds that
 \begin{eqnarray}\label{tra}
 E_{\rm out}=t_M \Big[1+\frac{2 g_{\epsilon}\overline{g}^\ast I(t) \sigma_z(0)}{\tilde{\gamma}+i(\tilde{\Delta}+\delta_{k_L})}\Big]E_{\rm in},
\end{eqnarray}
where $I(t)=1-\exp(-(\tilde{\gamma}+i(\Delta+\delta_{k_L}))t)$.
The angular overlap between $g_{\epsilon}$ and $\overline{g}^\ast$ can be evaluated using Eq.~(\ref{gave}) and Eq.~(\ref{geps}) and the sum rules for the spherical harmonics. We obtain
 \begin{eqnarray}
g_{\epsilon}\overline{g}^\ast&=&\epsilon(1- r_M e^{i\phi_0})\gamma,
 \end{eqnarray}
where $\phi_0=2k_0 R$, and $\gamma$ is the free space decay rate from the excited state.
Using contour integration for the Lamb shift, one finds
 \begin{eqnarray}\label{EQ29}
\tilde{\gamma}+i\tilde{\Delta}&=&(1- 2r_M\epsilon e^{i\phi_0})\gamma,
 \end{eqnarray}
such that
 \begin{eqnarray}
\frac{g_{\epsilon}\overline{g}^\ast}{\tilde{\gamma}+i(\tilde{\Delta}+\delta_{k_L})}&=& \frac{2\epsilon(1- r_M e^{i\phi_0}) }{1- 2r_M\epsilon e^{i\phi_0}+i\delta_{k_L}/\gamma},
\end{eqnarray}

where we discarded the diverging part of the Lamb shift in Eq.~(\ref{EQ29}), which
can be made finite by using a relativistic treatment and without making the rotating wave approximation \cite{het10}. By choosing
the driving laser to be resonant with the atomic transition ($\delta_{k_L}=0$, which would equal to the free-space Lamb-shifted transition frequency in a relativistic treatment), one can finally obtain
\begin{eqnarray}
T =  |t_M|^2  \Big|\frac{1- 2\epsilon }{1- 2r_M\epsilon e^{i\phi_L} }\Big|^2,
\end{eqnarray}
after averaging the equations over a state $|g,0\rangle$ and defining the steady state transmissivity as $T=\lim_{t \rightarrow \infty} \langle E_{\rm out}^{\dagger} E_{\rm out} \rangle/\langle E_{\rm in}^{\dagger} E_{\rm in}\rangle$.
It is the same result that we obtained by naively modeling
the atom as a mirror with reflectivity $2\epsilon$.
On resonance, the QED calculations yield the same results as the direct Fabry-P\'erot calculations.\\

\end{document}